\documentclass[fleqn,usenatbib]{mnras}
\usepackage{graphicx,amsmath,amssymb,url,xspace}
\usepackage{times}
\usepackage{bm}
\usepackage[usenames,dvipsnames]{color}
\usepackage{xfrac}

\newcommand{\Fig}[1]{Figure~\ref{#1}}
\newcommand{\Eq}[1]{Equation~(\ref{#1})}
\newcommand{\EQ}{\begin{equation}}
\newcommand{\EN}{\end{equation}}
\newcommand{\vv}{\mbox{\boldmath $v$} {}}

\newcommand{\DDD}{{\cal D} {}}
\newcommand{\uu}{\mbox{\boldmath $u$} {}}

\newcommand{\Sec}[1]{Section~\ref{#1}}

\newcommand{\ex}{{\bm{\hat{e}}_x}}

\newcommand{\nab}{\mbox{\boldmath $\nabla$} {}}

\newcommand{\St}{\text{St}}

\newcommand{\Orb}{\text{Orb}}
\newcommand{\grav}{\mbox{\boldmath $g$} {}}
\newcommand{\tdep}{t_{\text{gen}}}
\newcommand{\tdif}{t_{\text{diff}}}
\newcommand{\TTD}{\text{TTD}}
\newcommand{\cmcms}{\text{\,cm}^2\text{\,s}^{-1}}
\newcommand{\tR}{\tilde{R}}
\graphicspath{{.}{./Figures/}}

\date{Accepted XXX. Received YYY; in original form ZZZ}

\pubyear{2015}

\title[K abundance variations in the Solar Nebula]{Generating potassium abundance variations in the Solar Nebula}

\author[Alexander Hubbard]{Alexander Hubbard,$^1$\thanks{E-mail: ahubbard@amnh.org} \\
$^{1}$Department of Astrophysics, American Museum of Natural History, New York, NY 10024-5192, USA}

\begin{document}
\label{firstpage}
\pagerange{\pageref{firstpage}--\pageref{lastpage}}
\maketitle

\begin{abstract}
An intriguing aspect of chondritic meteorites is that they are complementary: while their separate components
have wildly varying abundances, bulk chondrites have nearly solar composition. 
This implies that
the nearly-solar reservoirs in which chondrites were born were in turn assembled from sub-reservoirs of differing
compositions that birthed the different components.
We focus on explaining the potassium abundance variations between chondrules even
within a single chondrite, while maintaining the observed CI $^{41}$K to $^{39}$K ratios.
{This requires physically separating potassium and chondrules while the temperature is high
enough for K to be in the gas phase.}
We {examine several mechanisms which could drive the dust through gas and}
show that to do so locally would have required long ({sub-orbital} to many orbits) time
scales;
{with shortest potassium depletion time scales occurring in a scenario where chondrules formed high above the midplane
and settled out of the evaporated potassium.}
While orbital time scales are at odds with laboratory chondrule cooling rate estimates, any other model for the
origin for the potassium abundance variation has to wrestle with the severe logistical difficulty of generating a plethora
of correlated reservoirs which varied strongly in their potassium abundances, but not in their potassium isotope ratios.

\end{abstract}


\section{Introduction}

Of all the planetary systems known, our own is the only one where we have detailed
elemental and isotopic data for any solid body. The dominant story told by that data is that,
excluding objects with exceptionally traumatic histories (e.g.~Mercury and the Moon), 
and focusing on the more refractory elements, solid objects throughout the solar system
have loosely comparable bulk abundances \citep{2000SSRv...92..237P}.
This is not surprising as protoplanetary disks
are expected to be quite {active}, and hence well mixed \citep{1973A&A....24..337S,1974MNRAS.168..603L}.
{
Indeed, even if there is a magnetic dead zone \citep{1996ApJ...457..355G,2014ApJ...791..137B}, turbulence from surface layers can penetrate and mix an otherwise
quiescent midplane \citep{2009ApJ...704.1239O}.}
Further, macroscopic solids in the solar nebula were made of vast numbers of sub-micron interstellar
grains, erasing any stochastic compositional variations between said grains \citep{2003ARA&A..41..241D}.

However, when examined in finer detail there are significant variations which point
to the solar nebula having contained separate reservoirs with different
abundances from which different classes of solids were formed \citep{2003TrGeo...2..547M}.
In particular for this paper, different chondrite classes
have modestly different abundances \citep{2006mess.book...19W}. Because chondrites are
undifferentiated, and sometimes only slightly thermally or aqueously altered, meteorites,
they provide an accessible record of the chemical and isotopic environments in which they and their components
formed \citep{1967GeCoA..31..747V}.

There is growing evidence for complementarity: chondrite components with strongly varying compositions
combine to form chondrites with overall solar bulk composition \citep{2010E&PSL.294...85H,2015E&PSL.411...11P}.
This implies that distinctly different parts of a chondrite, such as its chondrules and its matrix, were correlated with one another.
While complementarity between chondrules and matrix can be maintained in the face of turbulent mixing
for thousands of orbits assuming that the matrix
and chondrules originated from a single reservoir, it cannot be produced after the fact by
fortuitous radial transport \citep{2015arXiv150704009G}.
Complementarity therefore implies that the
separate larger scale reservoirs with mostly solar bulk composition from which the different chondrite classes
were assembled were, themselves, 
either divided into or assembled from smaller scale reservoirs with quite different, strongly non-solar, compositions.
In particular for this work, potassium abundances vary by factors
of many between different chondrules even within a given chondrite \citep{1991GeCoA..55..935H}.


{
As noted above, the sets of small scale reservoirs implied by meteoritic components of differing compositions
were correlated as is implied by complementarity.
If they were not themselves drawn in some fashion
from initial overarching large scale reservoirs of nearly solar composition which were preserved by the chondrite
assemblage process, then their very existence across multiple chondrite classes would require an awkward degree of fine tuning.}
As we will develop in greater detail, the differing potassium abundances imply the
existence of small scale reservoirs which were alternately
strongly depleted of, and weakly enriched in K. In this paper we examine the
conditions, and crucially the time scales, required for splitting an initially solar abundance reservoir into sub-reservoirs
with differing potassium abundances.
The results are generalizable and can be used to understand the lack of potassium isotope signatures throughout
the Solar System, even between objects with wildly varying bulk potassium abundances \citep{1995Metic..30Q.522H}.
The calculations further apply for generating any condensation temperature dependent depletions during planet
formation although the temperature histories implied would differ.

\section{Chondrules and chondrites}

Chondrites are a set of classes of undifferentiated meteorites composed of a mix
of chondrules, matrix, and other inclusions including Ca-, Al-rich inclusions (CAIs) \citep{2006mess.book...19W}.
Chondrules are sub-mm glassy clasts which were produced by melting dust
in the solar nebula, the protoplanetary disk in which our Solar System formed \citep{1997AREPS..25...61H}.
While the mechanism through which chondrules were made is as yet uncertain,
the melting required temperatures above $1700$\,K \citep{1990Metic..25..309H}.
Much of the space between chondrules
is filled with fine-grained matrix material which was not heated to such a degree. Indeed,
evidence suggests that only a small fraction of the matrix material could have been
heated above $700$\,K \citep{1995GeCoA..59..115H,2002GeCoA..66..661M} which means that models for
chondrule and chondrite formation
must allow for solids that experienced very different
thermal histories to have ended up in close proximity \citep{2015Icar..245...32H}.

Researchers have amassed a range of evidences arguing that different classes of 
chondrites sampled different chondrule or chondrule precursor
reservoirs \citep{2012M&PS...47.1176J}.
These include differences in oxygen isotope ratios, bulk compositions, chondrule sizes,
abundances of relic grains, and chondrule textural types.
While taken individually, these lines of evidence are not conclusive but combined they are persuasive enough that
in this paper we assume
that different chondrite classes did indeed form from distinct reservoirs. Note however that these distinct reservoirs
could have had identical bulk abundances and have
differed primarily how in the chondrules were
processed or how the chondrites were assembled: what matters for this paper is merely the logistical difficulty of generating
a multitude of large scale reservoirs which in turn contain a large number of correlated small scale reservoirs.

Among the classification schemes used for chondrules is one that depends on their olivine fayalite (Fe$_2$SiO$_4$) fraction: 
the more common Type I chondrules
have less than $10$ mol \% fayalitic olivine while the rarer Type II chondrules have Fa $>10$ mol \% \citep{1977GeCoA..41..477M,1977GeCoA..41.1843M}.
A further difference between the two types of chondrules is their moderate volatile content: Type I chondrules are strongly depleted in
moderately volatile elements such as sodium and potassium, while Type II chondrules are weakly enriched in those same 
elements 
\citep{1977GeCoA..41..477M,1977GeCoA..41.1843M,1989LPSC...19..523J,1990GeCoA..54.1785J}.
This potassium abundance difference will be our focus.

\section{Need for sub-reservoirs}
\label{sub_res_1}

Potassium is moderately volatile, with a $50\%$ condensation temperature of approximately
$1000$\,K under nebular conditions \citep{2003ApJ...591.1220L}. That is far lower than the chondrule melting temperature
of $1700$\,K, so
it is reasonable to imagine that differing thermal histories acting on two separate regions of the solar nebula
with identical bulk compositions led to the potassium depletion seen in Type I chondrules. In this paper we will show
that the time scales required for such processes are not insignificant.

Importantly, potassium has two abundant stable isotopes, $^{39}$K and $^{41}$K.
Therefore, straightforward partial evaporation would
Rayleigh fractionate those isotopes by preferentially evaporating the lighter $^{39}$K, leaving a heavy isotopic signature.
Neither heavy, nor indeed light, isotopic signatures have been found in chondrules for potassium, or for iron either
\citep{2000M&PS...35..859A,2001M&PS...36..419A}.
There are several broad categories of processes which could explain the
differing elemental abundances without leaving behind an isotopic signature.

Firstly, there are processes where the gas and dust were out of isotopic and elemental equilibrium with each
other when cooling through
the potassium condensation temperatures. In such cases Raleigh fractionation was unavoidable when and where
condensation occurred, but only to the degree said condensation actually occurred.
For Type I chondrules, that means that they either formed from
reservoirs which were tuned to result in both the observed elemental (potassium depleted with respect to CI) and isotopic (not 
potassium fractionated
with respect to CI) abundances, or that their chondrule
melting process must have been too short for meaningful evaporation and recondensation to have taken place.
Given the spread in the chemical signatures
seen across chondrules, it is implausible that so many different initial reservoirs existed whose
elemental and isotopic abundances were carefully tuned such that  their isotopic compositions converged upon melting, evaporation and
recondensation \citep{1991GeCoA..55..935H,1995Metic..30Q.522H}.

If on the other hand the heating and cooling associated with chondrule formation 
was so fast that no evaporation (and subsequent recondensation) occurred, then chondrules explicitly preserve the composition of their precursors.
In that case Type I and II chondrules still had to have sampled different
reservoirs, but the reservoirs would have only differed in their elemental,
but not isotopic, abundances.  In this case, the heating and cooling would have had to have been extremely
fast: more than $10000$\,K/hr \citep{1997LPI....28.1613Y,2000M&PS...35..859A,2004GeCoA..68.3943A},
while current models for chondrule formation struggle to reach
even $5000$\,K/hr \citep{2013ApJ...776..101B}. Further, experimental work has established that
such cooling rates are too hight to be consistent with the crystallization textures of most chondrules \citep[][and references therein]{2012M&PS...47.1139D}.

Secondly, the gas and dust could have been in equilibrium during at least the cooling phase of chondrule formation,
in which case there would be no isotopic signature as long as the total system never developed one
\citep{2005ASPC..341..432D,2008Sci...320.1617A}.
In equilibrium, as condensation occurs solids reabsorb evaporated species; so different equilibrium solid abundances
require different system abundances during cooling.
This case therefore requires reservoirs which differed in their elemental, but not isotopic, abundances.
We divide this case into two{: either equilibrium was maintained by not having evaporation and recondensation occur,
or it was maintained by having a sufficiently slow process.
It would have been possible to avoid evaporation and recondensation if the ambient gas had maintained exactly the correct amount of potassium (perhaps by
moving through a planetesimal
atmosphere \citealt{2012ApJ...752...27M}) to match the correct vapor pressure to avoid evaporation and condensation over the thermal history of a given chondrule and
for a range of observed potassium depletion patterns in chondrules. That scenario does not seem plausible. It would more naturally arise
if the solids were so concentrated that
only a negligible degree of K evaporation was needed to reach equilibrium K vapor pressures.}

{Such large (factors of more than $100$) increases in the dust concentration are difficult to achieve, especially for}
chondrule precursors which were presumably porous and therefore at least as well coupled to the gas as chondrules \citep{2012Icar..220..162J}.
While turbulence can 
preferentially concentrate same-sized dust grains \citep{2001ApJ...546..496C}, this requires
narrow dust size distributions \citep{2013MNRAS.432.1274H}.  The
meteoritic record shows modest but still significant scatter in chondrule sizes even within a given meteorite
\citep{Friedrich2014}, so large turbulent dust concentrations
are not expected to have occurred.  Worse, at the size and densities invoked the dust clouds would have been gravitationally
unstable \citep{2008Sci...320.1617A}, and gravitational collapse occurs quickly, on orbital time scales \citep{2009ApJ...704L..75J}.
If the chondrule formation mechanism
did not itself generate such dust concentrations, it would require excessive fine tuning for chondrule melting events to have
occurred only just before gravitational collapse at a rate sufficient to
explain chondrules, while leaving the matrix untouched.
Shock models for chondrule melting do concentrate the dust, but only by factors of order ten
\citep{2012ApJ...752...27M};
and even if such a coincidence of chondrule heating and dust concentration occurred, it
is difficult to imagine how the matrix, which was not heated and makes up a significant fraction
of even the ordinary chondrites, could have been evenly mixed in during the gravitational collapse
stage of planetesimal formation \citep{2015Icar..245...32H}.

In the cases of both rapid heating and cooling and extreme solid enrichment the initial dust composition matches
the final dust composition, so the origin of the reservoirs cannot be due to the chondrule formation process itself.
This merely pushes the origin of the Type I and II chondrule potassium difference back in time, and raises the question of
how different reservoirs with identical isotopic but different elemental abundances could have been generated without
invoking evaporation and condensation.  If the elemental composition differences
were inherited from pre-solar gas, the lack of an isotopic signature would be quite surprising; but any evaporation
and recondensation of K in the solar nebula would have required temperatures on
the order of $1000$\,K, high enough to have thermally processed matrix material. That has
been ruled out by laboratory analysis \citep{1995GeCoA..59..115H,2002GeCoA..66..661M}.

{Finally, we are left with} the gas and dust 
{having} cooled slowly through the condensation temperatures.
If they cooled slowly enough, isotopic equilibrium would have been maintained.
Further, local processes would have had time to split the large scale reservoir from which
a given chondrite was assembled into 
sub-reservoirs with differing total system (gas plus solids) potassium abundances.
In this case the chondrule formation process itself would generate the different potassium elemental abundances without
raising isotopic questions, greatly simplifying the logistical difficulties associated with generating many reservoirs
with varying elemental abundances.

\section{Generating sub-reservoirs}

\begin{figure}\begin{center}
\includegraphics[width=\columnwidth]{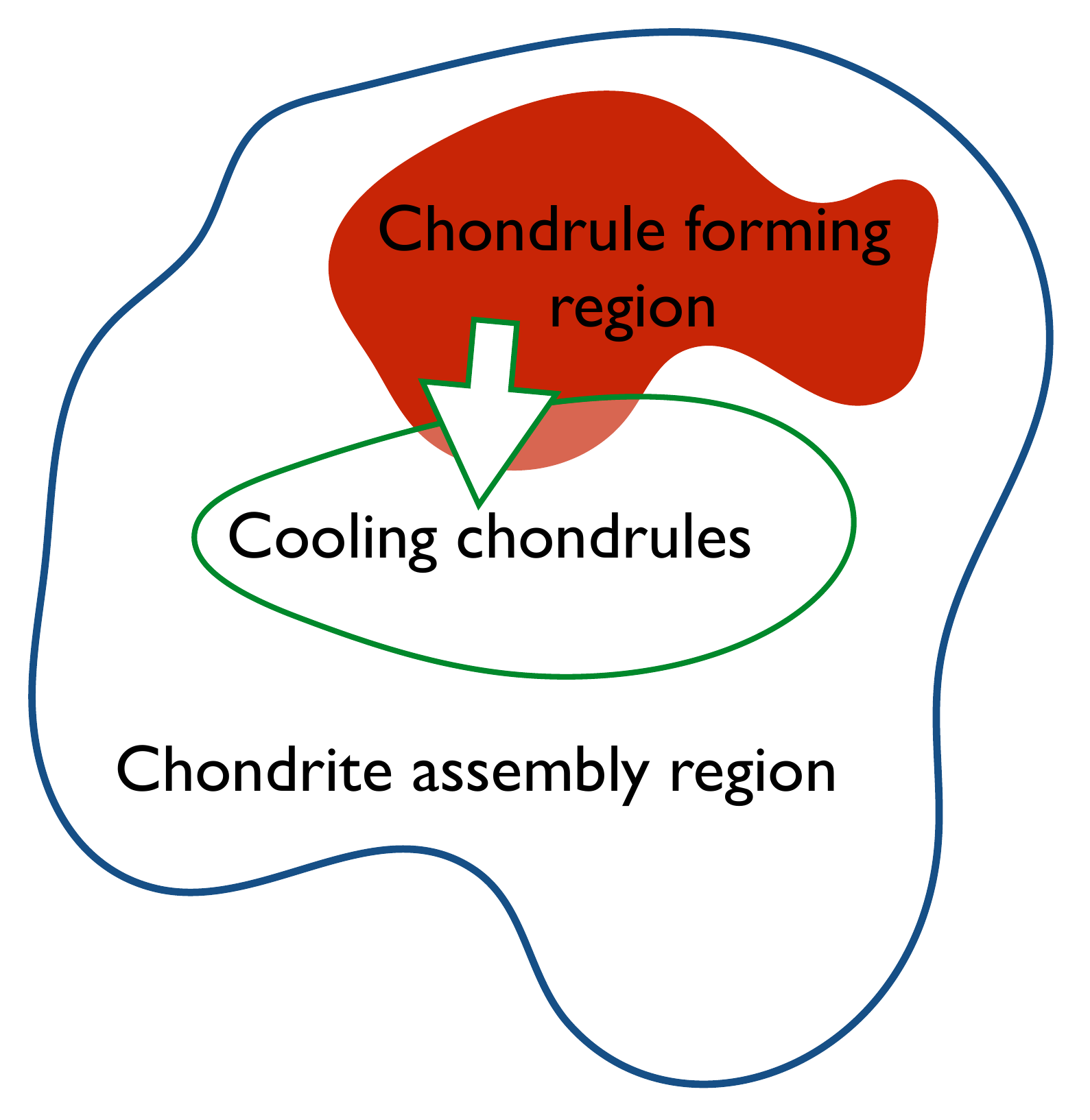}
\end{center}\caption{
Cartoon of how sub-reservoirs are generated within a larger chondrite assembly reservoir. The overarching reservoir maintains
complementarity. Chondrules move out of the chondrule forming region sufficiently slowly as to stay in K isotopic equilibrium with the gas.
The evaporation potassium does not move, so when the chondrules cool, they do so in a potassium depleted environment.
\label{Cartoon} }
\end{figure}

To enrich or deplete the eventual solids in a region of the solar nebula in a moderately volatile element such as K, that element needs to move
relative to the ``bulk'' less volatile elements. The bulk elements can be well represented by Si or Mg, which have similar
volatility ($50$\% condensation temperatures of order $1400$\,K),
{and which each provide approximately $20\%$ of the non-oxygen mass of a chondrule \citep{2003ApJ...591.1220L}.}
Note that because we are interested in the K/Si or K/Mg ratios, moving 
potassium out of a region
is equivalent to moving Si and Mg into that region, and vice-versa.
{At this stage we do not need to know the source of the drift speed, but only its strength.
We sketch a cartoon of this process in \Fig{Cartoon}. In it, we show a chondrite assembly region of CI composition.
A subsection of it heats to make chondrules, which then drift out of the heating area slowly, leaving
the evaporated potassium behind. When they finally cool, they do so in a region with only a small amount of potassium, so only
a small amount of potassium recondenses into the chondrules. Once that chondrule forming region also cools, most of the evaporated potassium would
instead recondense onto matrix grains.
Because the chondrules and the evaporated potassium stay within the larger chondrite assembly area, complementarity is maintained, and if the drift is slow enough,
isotopic equilibration would be maintained.}

{For simplicity, we assume that the relative motion occurs along a single direction, which we label with the coordinate $x$.}
If we assume that one participant (volatile or refractory) is tied to the nebular
gas and well mixed, while the other participant has a drift speed $\vv = v \ex$ and experiences a turbulent diffusion $\DDD$,
then a steady state is reached when the drift balances diffusion, i.e., when
\EQ
c v = \DDD \partial_x c, \label{diff_drift_balance}
\EN
where $c$ is the concentration (with respect to the nebular gas) of the mobile element. Solving \Eq{diff_drift_balance} we find
\EQ
c = c_0 e^{vx/\DDD}, \label{conc_z}
\EN
where $c_0$ is the concentration at $x=0$. \Eq{conc_z} leads us to define the length scale
\EQ
\ell_c \equiv \DDD/v.
\EN
Depletion or enrichment patterns then can be generated only on length scales $L > \ell_c$.
Diffusion is extremely effective at erasing small scale structures so
large drift velocities are required to maintain small scale structures, while
large scale structures are easier to preserve
in the face of diffusion.

However, steady states are not achieved instantly, and it will take a time $\tdep$ of order
\EQ
\tdep = L/v > \ell_c/v
\EN
to generate the depletions or enrichments: while large scale structures can achieve significant concentration
differences with only moderate drift speeds, those concentration differences require a long time to set up.

{
To make contact with standard disk theory we scale
\EQ
v= \beta c_s
\EN
and
\EQ
\DDD = \alpha c_s H
\EN
where $c_s$ is the gas background sound speed, $\beta$ measures $v$ in terms of $c_s$,
$\alpha$ is the Shakura-Sunyaev $\alpha$ parameter \citep{1973A&A....24..337S},  $H=c_s/\Omega_K$ the
gas scale height, and $\Omega_K$ is the Keplerian frequency.}
Note that we will use $c_s$ to refer
to the background disk sound speeds; and $u_{th}$ for the gas thermal speed in the potassium condensation regions:
the existence of cold matrix implies that most of the solar nebula remained cold, well below
K condensation temperatures so chondrule forming regions are expected to contain only a small fraction
of the total volume \citep{2015Icar..245...32H}.

With those scalings, we have
\EQ
\frac{\tdep}{\Orb} = \frac{\DDD}{v^2 \Orb} =  \frac{\alpha}{2 \pi \beta^2}, \label{Eq_t_Orb}
\EN
where $\Orb \equiv 2\pi/\Omega_K$ is the local orbital period. Note
that $\alpha$ disks with turbulence driven by orbital shear assume turbulent speeds of order
\EQ
u_t \sim \sqrt{\alpha} c_s.
\EN
If the drift speed is due to turbulence, we would expect $v \lesssim u_t$ and so
$\beta \lesssim \sqrt{\alpha}$ and $\tdep/\Orb \gtrsim 1/2 \pi$. 

\begin{figure}\begin{center}
\includegraphics[width=\columnwidth]{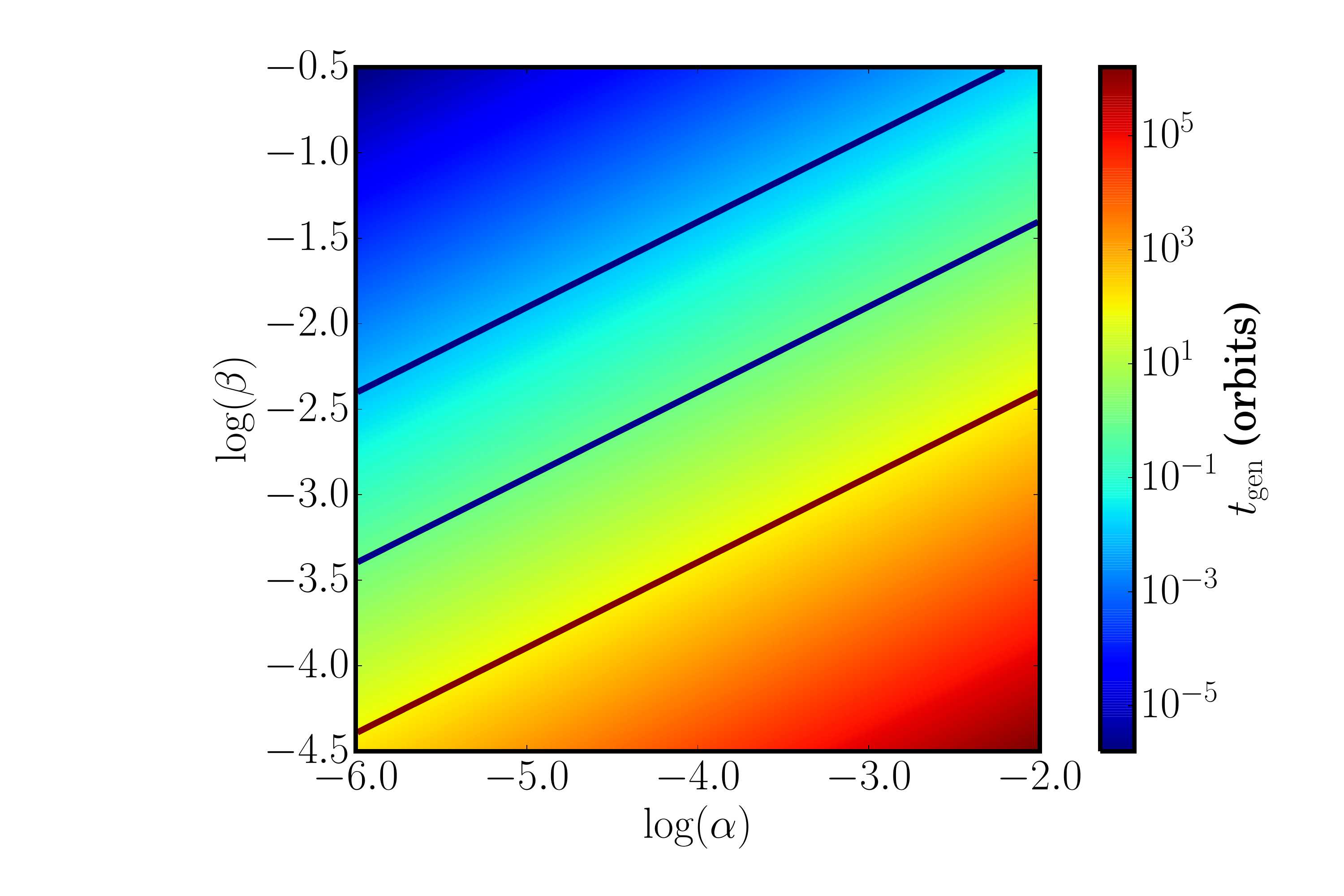}
\end{center}\caption{
Timescale $\tdep$ in orbits as a function of $\alpha$ and $\beta$ (Equation~\ref{Eq_t_Orb}). From top to bottom the lines
are $\tdep=10^{-2}, 1, 100$\,Orb.
\label{times} }
\end{figure}

We plot \Eq{Eq_t_Orb} in \Fig{times}, showing curves of $\tdep= 10^{-2}, 1, 100$\,orbits. We use $\alpha=10^{-6}$ as a lower
cut-off for the turbulent diffusion because even very laminar disks
are expected to be somewhat turbulent and said turbulence will propagate from
the turbulent surface layers to the midplane \citep{2009ApJ...704.1239O}.
{Note that to manage $\tdep<100$\,Orb for $\alpha=10^{-6}$ we require $\beta>4 \times 10^{-5}$.}

\section{Drift Speeds}
\label{drift_speeds}

As discussed in \Sec{sub_res_1}, the different reservoirs that produced 
Type I and II chondrules respectively must have existed within the already different reservoirs that produced the different chondrite classes.
To avoid a fine tuning problem we have adopted the assumption that chondrule formation itself naturally produced sub-reservoirs with differing levels of potassium.
This process must occurred slowly enough that K isotopic equilibration
between gas and dust was continuously maintained.  We will also assume elemental equilibration between gas and solids, i.e.~that 
the composition of the solids is only a function
of the total (gas and solids) composition and the temperature. 
This assumption only applies during chondrule formation and any subsequent thermal processing; 
and is relaxed for temperatures well below $1000$\,K, when
equilibration would take such long time scales that compositions can be considered frozen.

Generating such sub-reservoirs required moving K with respect to Si and Mg, without moving potassium isotopes
with respect to one another. The mobile element(s) could have been in the solid or gas phase.
If all the elements were in the solid phase, this requires both that there existed categories of dust rich in and depleted in
potassium, and that these categories of dust had
different aerodynamical properties.  We neglect that possibility because
this merely pushes the sub-reservoir problem back to the dust growth stage. Note that a $0.25$\,mm \citep{Friedrich2014} chondrule
made from $0.1$\,$\mu$m interstellar grains would contain the record of $15 \times 10^9$ of them, while chondrule precursors would have have K mass fractions of order
 $5 \times 10^{-4}$ \citep{2003ApJ...591.1220L}. This means that K is far too abundant for nugget effects to explain its abundances variations between
chondrule sized samples.

{
In what follows we will assume a Hayashi Minimum Mass Solar Nebula \citep[MMSN,][]{1981PThPS..70...35H} normalized to $R=2.5$\,AU, the location of the asteroid
belt:
\begin{align}
&T_g \simeq 177 \tR^{-1/2} \text{K}, \\
&c_s \simeq  8 \times 10^{4} \tR^{-1/4} \text{ cm s}^{-1},  \label{cs} \\
&\Omega_K \simeq 5 \times 10^{-8} \tR^{-3/2} \text{s}^{-1}, \\
&\Sigma_g \simeq 430 \tR^{-3/2} \text{ g cm}^{-2}, \label{Sigmag} \\
&\rho_0 \simeq 10^{-10} \tR^{-11/4} \text{ g cm}^{-3}, \label{rhog0}
\end{align}
where $\tR \equiv R/2.5\,\text{AU}$.
Unless otherwise specified, we will assume
that the local (hot) gas had a temperature $T_g=1000$\,K with associated thermal speed 
\EQ
u_{th} = \sqrt{\frac{2 k_B T}{m}} \simeq 2.7 \times 10^5 \text{ cm s}^{-1}, \label{uth}
\EN
where 
\EQ
m= \mu m_H
\EN
was the mean molecular mass of the gas ($\mu\simeq 2.33$).
Combining Equations~(\ref{cs}) and (\ref{uth}), we find
\EQ
u_{th}/c_s \simeq 3 \tR^{1/4}.
\EN
}

{
Because we are interested
only in the potassium signal, and can take chondrule formation as a laboratory given,
the details of the dust dynamics prior to chondrule formation, and the details of the chondrule formation
scenarios themselves are not vital. Different chondrule formation scenarios naturally allow for different drift speeds, and requirements
on $\tdep$ will limit the space of viable chondrule formation scenarios.}

\subsection{Gas motion}

If two species were in the gas phase, then for one to have moved with respect to another at meaningful speeds
there needs to have existed a force acting differently on two species.
An atom of a given species $i$ of mass $m_i$ will encounter its own mass in other atoms in a time
\EQ
t_i = \frac{m_i}{n \sigma u_{th}  m},
\EN
where
\EQ
n = \frac{\rho_g}{\mu m_H},
\EN
is number density of the gas and
$\sigma$  is the effective collisional cross-section.

A force density $f$ exerted on species $i$ (and only on species $i$) would then drive a drift velocity
\EQ
u_i \sim \frac{f}{\rho_i} \times t_i \simeq \frac{f  m_i}{\rho_i \rho_g \sigma u_{th}}.
\label{v_0}
\EN
Most forces in protoplanetary disks act on the bulk fluid (e.g.~thermal pressure and gravity), 
and would only drive drift due to smaller differential velocities between different species with differing
masses and collisional cross-sections.
However, magnetic pressure offers a possible force $f$
which could act on potassium differently than on Mg or Si.  

{
Neutral particle do not feel magnetic forces, so magnetic pressure acts
on the disk's gas by exerting a force on charged ions which in turn collide with the mostly neutral gas.
Because the ion/neutral coupling is not quite perfect, the ions undergo a slow
ambipolar drift through the gas according to \Eq{v_0}.
Chondrule forming regions are hot enough
to thermally ionize potassium.}
If K$^+$ is the dominant ion, then the potassium will
experience that drift. Note however that type I and II chondrules also differ in their sodium abundances,
and sodium is significantly harder to thermally ionize than potassium, {and so would drift less.}

We can place limits on $f$ by comparing it to
\EQ
f_1 \equiv \frac{\rho_g c_s^2}{H},
\EN
the force density required to maintain
a $100\%$ bulk gas density variation over a local gas scale height. For $f=f_1$, we find
\EQ
u_i \simeq\frac{  m_i \Omega_K}{\rho_i  \sigma} \frac{c_s}{u_{th}}.
\label{v_1}
\EN
In that case, and for $\sigma \gtrsim 10^{-14}$\,cm$^2$ and
$\rho_K/\rho_g \simeq 4 \times 10^{-6}$, we have
\EQ
v_K \simeq 0.3\,\text{cm s}^{-1},
\EN
{
which corresponds to
\EQ
\beta \simeq 3 \times 10^{-6} \tR^{1/4}.
\EN
That value for $\beta$ is too low to be included on \Fig{times}, and thanks to the weak radial dependence requires $\tdep \gg 100$\,Orb at any orbital position for which
the MMSN model applies.
}

Even in the presence of an extreme magnetic pressure force, and with only extremely weak turbulence,
generating elemental abundance differences purely in the gas phase requires many century-scale times.
In practice magnetic pressures are lower than assumed in this calculation, and other ions also bear part of
the burden (which is important because if K is the only species providing ions, then only the K abundances
would be altered through ambipolar drift), both of which would
act to reduce $v_K$ and increase $\tdep$ even further. 

\subsection{Dust motion}

At temperatures between about $1400$\,K $ > T > 1000$\,K,
where potassium is partially or fully in the gas phase while Si and Mg are
fully condensed, moving the solids through the gas will alter the K/Si and K/Mg ratios.
The motion will not inherently generate
an isotopic signature because the motion of a dust grain does not meaningfully depend on the difference
between even a significant fraction of $^{39}$K or $^{41}$K atoms.  However, if the gas and dust are not in isotopic equilibrium,
then dust motion will transport one isotope preferentially.  We must therefore assume long enough time scales
for the solid motion that equilibration between gas and dust occurs continuously
(for a similar process, see \citealt{2014Icar..237...84H}).  Further, to reproduce
the degree of potassium depletion seen in Type I chondrules ($\sim 80\%$, \citealt{1991GeCoA..55..935H}), the temperature
must have been elevated enough for K to have mostly evaporated while the dust was in motion with
respect to the gas.

\subsubsection{Settling}

Dust embedded in gas tends to drift towards pressure maxima \citep{1977MNRAS.180...57W}.
The strongest non-transient pressure gradient in
a protoplanetary disk is the vertical one: residual gravity pulls gas and dust towards the midplane, and while
a vertical pressure gradient can maintain the gas in hydrostatic equilibrium, the dust does not feel the
pressure forces and settles downwards.
{Other pressure maxima exist in disks, such as zonal flows \citep{0004-637X-763-2-117}, although their strength is comparatively limited and
they are not expected to last for the entire disk life time. As such, barring extreme examples which are unlikely to have been correlated
with chondrule formation (perhaps a radial
gap opened by a planet, \citealt{2012ApJ...755....6Z}), settling can be taken
as the strongest drift speed due to gas pressure gradients.}

Dust motion through gas is controlled by drag. The equation of motion for dust is
\EQ
\partial_t \vv_d = -\frac{\vv_d-\uu_g}{\tau_d} + \grav +  \cdots,
\EN
where $\vv_d$ is the dust velocity, $\uu_g$ the gas velocity at the dust grain's position, $\grav$ the gravitational
acceleration and $\tau_d$ the drag time.

{Chondrules have radii significantly smaller than the expected gas molecular mean free path (tens of centimeters
for $\rho_g = 10^{-10}$g cm$^{-3}$), which places them in the Epstein drag regime,} with
\EQ
\tau_d = \frac{a \rho_d}{u_{th} \rho_g},\label{t_d}
\EN
where $a$ and $\rho_d$ are the dust grain radius and solid density, and
$u_{th}$ is the actual local gas thermal speed rather than the background disk sound speed $c_s$. We non-dimensionalize $\tau_d$ through
\EQ
\St \equiv \tau_d \Omega_K,
\EN
the dust Stokes number. 

Noting that for a gas disk in vertical hydrostatic equilibrium
\begin{align} 
&\rho_g = \rho_0 e^{-z^2/2H^2}, \\
& \rho_0 = \frac{\Sigma_g}{\sqrt{2 \pi} H},
\end{align}
where $\rho_0$ is the midplane gas density, we can write
\EQ
\St = \tau_d \Omega_K =\left( \frac{c_s}{u_{th}} e^{z^2/2H^2}\right) \St_0, \label{Eq_St}
\EN 
where
\EQ
\St_0 \equiv \frac{a \rho_d}{c_s \rho_0} = \frac{\sqrt{2 \pi} a \rho_d}{\Sigma_g} \label{St0}
\EN
is the dust Stokes number at the midplane.
{
Combining Equations~(\ref{Sigmag}) and (\ref{St0}) we find that
an $a=0.25$\,mm, $\rho_d = 3$\,g\,cm$^{-3}$ chondrule would have had
\EQ
\St_0 = 4 \times 10^{-4} \tR^{3/2}. \label{St_est}
\EN
For a dust grain of radius $a$ at constant mass $m_d$,}
\EQ
a = \left(\frac{3 m_d}{4 \pi \rho_d}\right)^{1/3} \propto \rho_d^{-1/3}.
\EN
From \Eq{t_d} we can
see that at constant mass then,
\EQ
\St \propto a \rho_d \propto  \rho_d^{2/3}, \label{St_porous}
\EN
and note that more porous particles are better coupled to the gas.
{The porosity of collisionally grown aggregates is as yet uncertain, but in the absence of
thermal processing is certainly significant ($80-95$\%) \citep{2010A&A...513A..57Z} and if compaction
is inefficient, could be
almost arbitrarily close to unity \citep{2008ApJ...677.1296W}.}

From \Eq{St_est}, we can see that chondrule-sized grains are small enough that terminal velocity
can be assumed:
\EQ
v_d = |g \tau_d|.
\EN
The vertical force of gravity in disks is
\EQ
g_z = -z \Omega_K^2,
\EN
which results in a settling speed
\EQ
v_s = \St\, z\, \Omega_K,
\EN
{Using $H = c_s/\Omega_K$ and \Eq{Eq_St} we arrive at
\EQ
\beta \equiv  \frac{v_s}{c_s} = \St \frac zH =\left(\St_0 \frac{c_s}{u_{th}}\right) \left(\frac zH  e^{z^2/2H^2} \right). \label{beta_settling}
\EN
Equations}~(\ref{St_porous}) and (\ref{beta_settling}) show how chondrule formation events would be correlated with
settling: thermal processing of high porosity dust leads to compaction, reducing drag, and raising $\St$
\citep{2014Icar..237...84H}.

\begin{figure}\begin{center}
\includegraphics[width=\columnwidth]{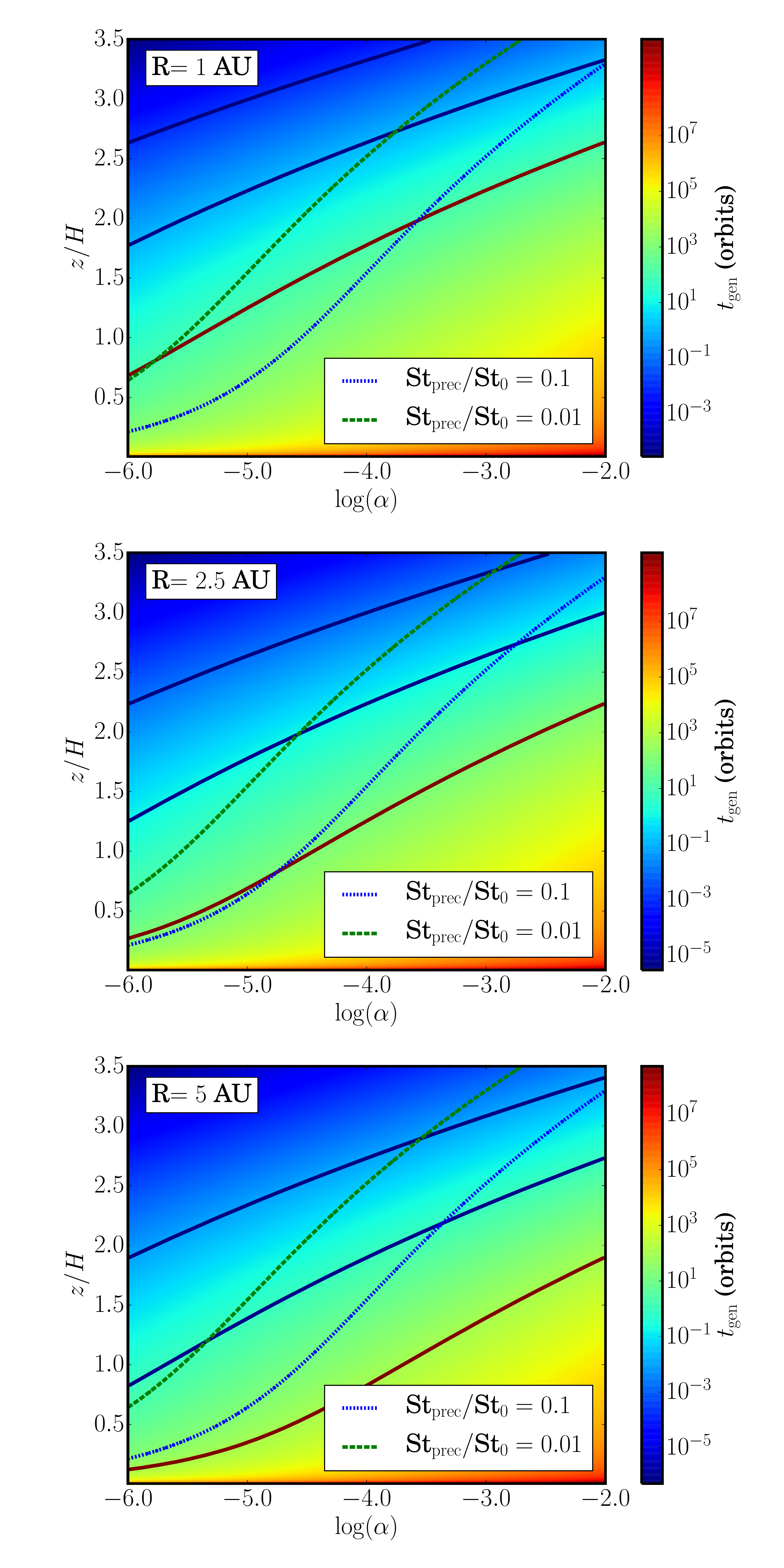}
\end{center}\caption{
Timescale $\tdep$ in local orbits assuming that $v$ is due to settling, for $R=1, 2.5,$ and $5$\,AU as a function of $z/H$. From top to
bottom the solid lines are $\tdep=10^{-2}, 1, 100$\,Orb. The dotted and dashed lines are the highest altitude to which turbulence
will mix dust with $\St_{\text{prec}} = 0.1, 0.01\,\St_0$, respectively.
\label{times_settle} }
\end{figure}

{
Estimating
\EQ
\St_0 \frac{c_s}{u_{th}} \simeq 1.5 \times 10^{-4} \tR^{5/4}, \label{St0_est}
\EN
we can use Equations~(\ref{Eq_t_Orb}) and (\ref{beta_settling}) to
calculate $\tdep$ as a function of $z/H$ and $\alpha$ as shown in \Fig{times_settle} for $R=1, 2.5,$ and $5$\,AU.
As might be expected, the lower gas densities at larger orbital positions leads to settling being faster.}
 While $\tdep$ could be less than
$0.01$\,Orb for low levels of turbulence 
at modest altitude, such weak turbulence would be unable to strongly loft the chondrule precursors.
\cite{2002ApJ...581.1344T} found that the maximum height $h$ to which turbulence will loft grains is approximately
\EQ
\frac hH = \sqrt{2 \ln\left(1+\frac{\alpha}{\St}\right)}.
\EN
Chondrule precursor grains with initial porosities of $97, 99.9\%$ would have had
$\St_{\text{prec}}=0.1, 0.01\,\St_0$. The dotted and dashed lines in \Fig{times_settle} show the maximum height
to which such precursors grains could have been lofted.

{
In this scenario, values of $\tdep<0.01$\,Orb were therefore possible only
for extremely porous precursor grains (porosities above $99.9\%$) at very high altitudes ($z/H>3$) at significant ($R>2.5$\,AU) orbital positions,
although the altitude and porosity constraints would relax for very large $R$.
Even values of $\tdep =1$\,Orb still require
highly ($\sim 97\%$) porous grains and high ($z/H \sim 2.5$) altitudes at $R=2.5$\,AU although again, those constraints
are somewhat relaxed by $R=5$\,AU.}
Note that only about $1.25\%$ of the gas mass
was at heights above $z=2.5$H, and even less of the partially settled, dust mass,
so chondrule formation would need to have been quite efficient if restricted to that altitude.

{While \Fig{times_settle} shows that settling is more effective at larger orbital positions, it measure $\tdep$ in units of the local
orbit, which also increases with $R$. While measuring $\tdep$ in orbits is useful for comparisons to disk dynamical (e.g.~turbulent)
time scales, it is less useful for comparison to chemical time scales.
For those purposes, we invoke \Eq{St0_est} to write
\EQ
\tdep = \frac{\alpha}{\beta^2 \Omega_K} \propto \tR^{-1},
\EN 
so for settling, larger $R$ is associated with shorter absolute time scales as well.
}

\subsubsection{Photophoresis}

Photophoresis is a process in
which differences in the illumination on two sides of a dust grain generates and maintains a temperature gradient through
the grain \citep{1918AnP...361...81E}.  As a result, gas molecules rebound faster from the hot side than the cold side, 
creating a net
force that pushes the dust towards its cold side.
{Photophoresis depends strongly on the details of dust grains' internal thermal conduction.
While that can become extremely involved for porous grains \citep{2016MNRAS.455.2582M},
chondrule forming regions were by definition hot
enough to melt the dust, leaving only comparatively simple solid silicate spheres.}

{To trap their heat, chondrule forming regions were presumably optically thick, and unable to feel stellar irradiation.
However, they were also hot enough to generate their own strong radiation fields that could have driven photophoresis.}
Recently \cite{2015ApJ...814...37M} derived the photophoretic velocity in the optically thick limit, finding that for solid chondrules  
with thermal diffusivity $k \simeq 1.5 \times 10^5$\,erg\,s$^{-1}$\,cm$^{-1}$\,K$^{-1}$,
\EQ
v_p  \simeq 2.28 \times 10^{-8} \Gamma \left(\frac {T}{\text{K}}\right)^{7/2} \text{cm s}^{-1}, \label{vp}
\EN
where $\kappa_R$ is the Rosseland mean opacity and
\EQ
\Gamma \equiv -\frac{1}{\kappa_R \rho_g} \frac{\partial \ln T_g}{\partial z}
\EN
measures the temperature gradient in units of optical depths.
\cite{2014ApJ...791...62M} found $\Gamma \sim 10^{-4} - 10^{-3}$ for MRI turbulence, although those models
assumed a temperature independent opacity and larger values could easily occur under more realistic assumptions,
and could have driven much larger values of $v_p$.

\begin{figure}\begin{center}
\includegraphics[width=\columnwidth]{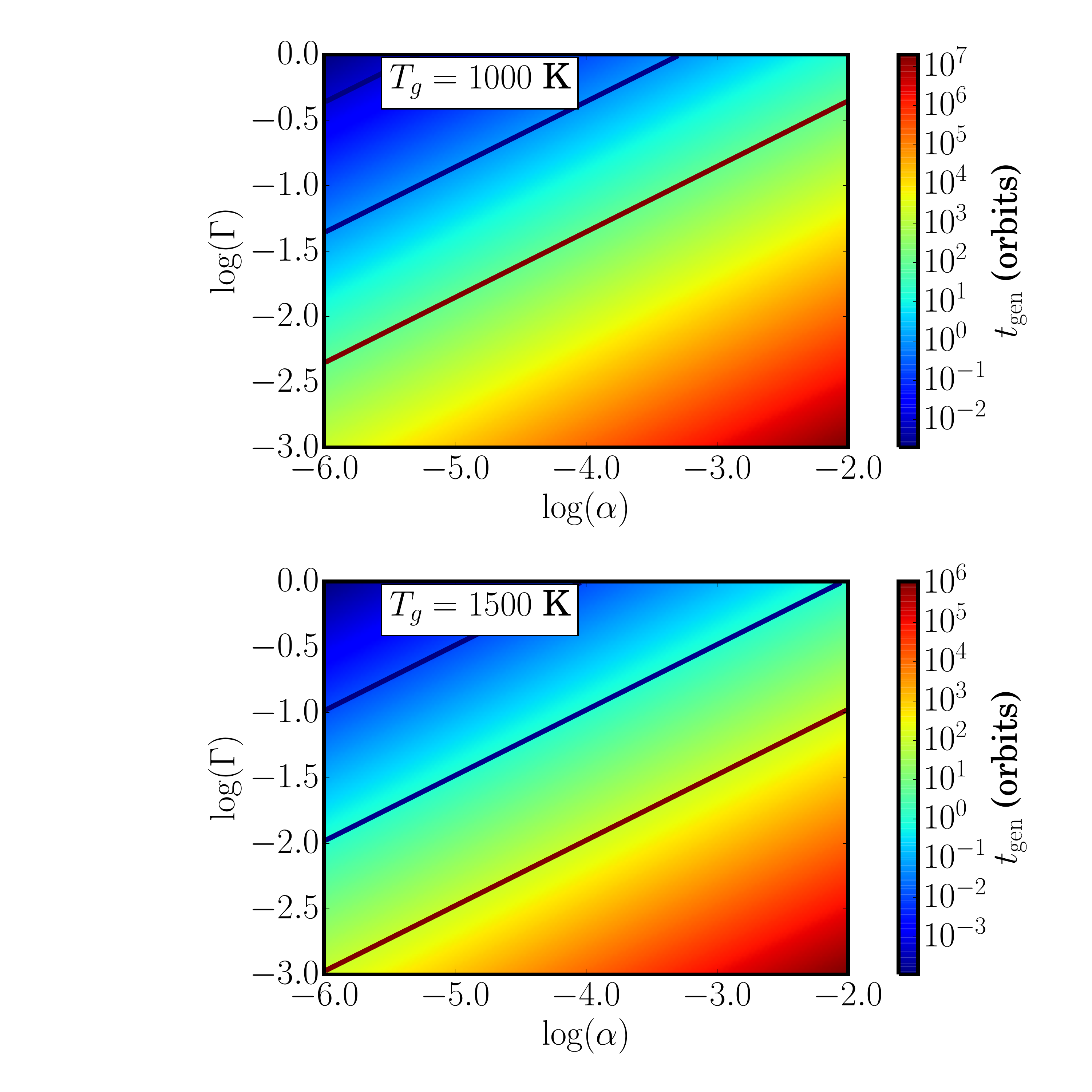}
\end{center}\caption{
Timescale $\tdep$ in orbits at $R=2.5$\,AU, assuming that $v$ is due to photophoresis, for $a=0.25$\,mm fused silica grains.
Top panel: $T_g=1000$\,K. Bottom panel: $T_g = 1500$\,K.
From top to
bottom the solid lines are $\tdep=10^{-2}, 1, 100$\,Orb. 
\label{times_photo} }
\end{figure}

{
In \Fig{times_photo} we show the time scale $\tdep$ as a function of $\alpha$ and $\Gamma$ for $T_g=1000, 1500$\,K
at $R=2.5$\,AU. Values of $\tdep<0.01$\,Orb
were possible only for extremely large ($\Gamma>0.1$) temperature gradients combined with the weakest turbulence
we consider ($\alpha \sim 10^{-6}$).}
Values of $\tdep=1$\,Orb are more plausible, but still required significantly
larger values of $\Gamma$ than have been seen to date combined with extremely laminar disks.
{
In \Eq{vp}, $T$ is due to the chondrule forming event, rather than the background disk temperature,
so $v_p$ is independent of orbital position. From \Eq{Eq_t_Orb}, we therefore have
\EQ
\frac{\tdep}{\Orb} \propto c_s^2 \propto \tR^{-1/2},
\EN
and 
\EQ
\tdep \propto \tR.
\EN
In terms of dynamical time scales, photophoresis is more effective at larger $R$, but in terms of absolute time scales, it
is more effective at smaller $R$. Neither of the dependencies is strong enough to alter the above conclusions for reasonable
ranges of $R$.
}

\subsubsection{Turbulent thermal diffusion}

Turbulent thermal diffusion (TTD) is a process where turbulence preferentially transports moderately well coupled dust grains down gas
temperature gradients (from hot to cold) \citep{1996PhRvL..76..224E}.
A quick reading of \cite{2015arXiv151202538H} would suggest that the regions we are considering are warm enough
that thermal relaxation would halt  the TTD: if turbulent eddies rapidly thermally equilibrate with their surroundings the effects of
the temperature gradient are reduced.
However, regions associated with chondrule formation are often
quite rarified. A $R=2.5$\,AU MMSN midplane at $177$\,K would have a gas density of $\rho_g = 10^{-10}$\,g\,cm$^{-3}$,
and higher temperature or altitude regions would be expected to be even less dense. Further,
during chondrule formation we expect nebular fines to evaporate, and large grains to compactify,
resulting in relatively poor gas-dust thermal coupling, and the gas-dust thermal coupling time can dominate over turbulent time scales.

In high temperature regions, without ices,
the gas-to-dust ratio for a mixture of nebular composition is approximately $200$ \citep{2003ApJ...591.1220L}. Accordingly, a single grain of mass $m_d$
mediates the radiative thermal relaxation for a parcel of gas of mass $200 m_d$. A temperature perturbation of $\triangle T$ for that
parcel corresponds to a thermal
energy perturbation of
\EQ
\triangle e = 200 m_d \times \frac{k_B T}{\left(\gamma-1\right) m},
\EN
where $\gamma\sim 1.4$ is the adiabatic index. Even if the dust temperature is set by radiative equilibrium with the surroundings, the gas will take a time
\EQ
t_{th} = \frac{\triangle e}{4 \pi a^2  u_{th} \rho_g \alpha_{\text{tac}}  \frac{k_B T}{(\gamma-1) m}} \label{t_dust_gas_0}
\EN 
to thermally equilibrate with the dust,
where the factor of $4\pi a^2 u_{th} \rho_g$ is rate at which gas impinges on the dust grain, and
$\alpha_{\text{tac}}$ is the thermal accommodation coefficient.
\Eq{t_dust_gas_0} can be expanded to find
\EQ
t_{th} \simeq 2 \times 10^5  \left(\frac{u_{th}}{2.7 \times 10^5 \frac{\text{cm}}{\text{s}}}\right)^{-1} \left(\frac{\rho_g}{10^{-10} \frac{\text{g}}{\text{cm}}}\right)^{-1} \text{s}.
\EN
{
This implies
\EQ
t_{th} \Omega_K \simeq 10^{-2} \tR^{5/4}. \label{tth}
\EN
}

From appendix A6 of \citep{2015arXiv151202538H}, the turbulent thermal diffusion velocity is
\EQ
v_{\text{TTD}} \simeq  -\frac 89 C \tau_s \frac{k_B T}{m} \ln\frac{k_1}{k_2} \partial_z T, \label{VTTD0}
\EN
where $k_1$ and $k_2$ are the limiting wavenumbers of the turbulence which contributes to the TTD, and $C \simeq 0.3$ is a constant
of order unity. The upper limit, $k_1$
is set by the stopping time of the dust grains: turbulent motions with shorter correlation times than the dust stopping time
do not contribute. The thermal relaxation time sets the lower limit $k_2$: turbulent motion with correlation times longer than the thermal relaxation
time do not contribute.
For a Kolmogorov cascade, the wavenumber scales with the turbulent correlation time as
\EQ
k \propto t^{-3/2},
\EN
so \Eq{VTTD0} becomes
\EQ
v_{\text{TTD}} \simeq  -\frac 43 C \tau_s \frac{k_B T}{m} \ln\frac{t_{th}}{\tau_s} \partial_z T. \label{VTTD1}
\EN
We can estimate
\EQ
\ln\frac{t_{th}}{\tau_s} \simeq 3
\EN
using Equations~(\ref{St_est}) and (\ref{tth}).

{
If we further assume that the temperature gradient was comparable to the local
scale height, then $|\nab \ln T_g| = H^{-1}$, and at $R=2.5$\,AU, 
\EQ
v_{\TTD} \sim 216 \text{ cm s}^{-1}, \label{VTTD2}
\EN
which corresponds to
\EQ
\beta =2.7 \times 10^{-3}.
\EN
That value of $\beta$ would have been sufficient to allow for short $\tdep < 1$\,Orb for weak, $\alpha<5 \times 10^{-5}$, turbulence.}
However, \Eq{VTTD0} does not apply when $v_{\TTD}$ would be comparable to the turbulent speed $u_t$ at the length scale $k_0$.
Assuming that the largest scale turbulence has a velocity scale $\sqrt{\alpha} c_s$ and a time scale $\Omega_K^{-1}$, we have
\EQ
u_t (k_0) \simeq \sqrt{\alpha} c_s k_2^{-1/3} \simeq \sqrt{\alpha t_{th} \Omega_K} c_s.
\EN 
{
We therefore have the further constraint that $v_{\TTD} <  \sqrt{\alpha t_{th} \Omega_K} c_s$,
which implies $\beta < \sqrt{\alpha t_{th} \Omega_K}$ and hence 
\EQ
\frac{\tdep}{\Orb}  \gtrsim  \frac{\alpha}{2 \pi \left(\alpha t_{th} \Omega_K\right)} \simeq 16 \tR^{-5/4},
\EN
and
\EQ
\tdep \propto \tR^{1/4}.
\EN
}

\section{Comparison to cooling constraints}

As shown in \Sec{drift_speeds}, for reasonable ranges of $R$, values of $\tdep<0.01$\,Orb were only achievable through settling, and even then,
only if chondrule formation were limited to extreme altitudes ($z/H>3$) and chondrule precursors were extremely porous (less than
$10^{-3}$ volume filling fraction). Values of $\tdep$ on the order of an orbit was possible through settling at more reasonable, if still high altitudes
($z/H>2.5$) and for significantly more compacted precursors ($97\%$ porosities). 
At the midplane, turbulent thermal diffusion can drive
$\tdep \sim 20\, \Orb$, but significantly smaller values are not expected except for very small $R$.
On the other hand, we do not expect
 hot thermal structures in the disk to have been extremely long lived, so $\tdep<100$\,Orb would seem a reasonable upper limit.
 
Accordingly,
generating chondrules with significant potassium depletion through evaporation required timescales
\EQ
0.01\,\text{Orb} < \tdep < 100\,\text{Orb}. \label{t_const}
\EN
If those time scales were not available or are otherwise ruled out, the origin of the potassium abundance variation between chondrules
becomes a significant logistical problem.
While the details do depend on the orbital position, the dependencies are not extreme, and these conclusions easily hold for $1$\,AU $< R<5$\,AU.

\subsection{Chondrule cooling times}

These time scales required in \Eq{t_const} are in significant tension with laboratory chondrule cooling constraints.
{While the experimental limits on chondrule cooling rates
vary significantly:  $10-1000$\,K/hr \citep[][and references therein]{2012M&PS...47.1139D}.}
Even a very low rate of $10$\,K$/$hr would cool from $1700$\,K to $1000$\,K in under three days, approximately
$t_{\text{cool}} \lesssim 2\times 10^{-3}$\,Orb at $R=2.5$\,AU; {and the upper end of the cooling rate are far faster, with corresponding
far shorter $t_{\text{cool}}$.} The laboratory
constrained cooling time scales put upper limits on the cooling time which are far too short to have allowed particles time
to move through the gas and set up potassium abundance variations (i.e.~$\tdep > 0.01$\,Orb $\gg t_{\text{cool}}$).

However, these cooling times $t_{\text{cool}}$ are not strictly incompatible with $\tdep  \gg t_{\text{cool}}$
because the cooling rate limits primarily measure thermal histories at crystallization temperatures well above the K condensation temperature
which constrains $\tdep$. The two constraints can be simultaneously satisfied if the chondrules cooled rapidly through their liquidus
temperatures \citep[$T \gtrsim 1400$\,K,][]{1990Metic..25..309H}, matching observed crystallization patterns;
but then lingered for orbital time scales at K recondensation temperatures ($\sim 1000$\,K) allowing the chondrules to move
sufficiently through the gas to generate the observed potassium abundance variations.

\section{Isotopic equilibrium and zoning}

A further constraint comes from the lack of a potassium isotopic signature.
In the above analysis leading to \Eq{t_const} we have only calculated the time scale required to move chondrules
to low potassium regions while the K is entirely evaporated. This means that we have assumed
that $\tilde{P}_K \ll P_{K, sat}$, where $\tilde{P}_K$ is potassium's partial pressure were it entirely in the gas phase,
and $P_{K, sat}$ its saturation vapor pressure.
However, when the potassium recondensed the system moved to the opposite limit, with $\tilde{P}_{K} \gg P_{K, sat}$.
This transition was presumably due to cooling, either because the chondrules
moved to regions of lower ambient temperature, or the entire region cooled.

Note that once $K$ began to recondense, the chondrules contributed to the net transport of potassium; and it
is imaginable that, were they to have drifted sufficiently rapidly through the gas, they could have entered a region where $K$ would evaporate.
We will therefore assume that, especially given the long $\tdep$ we predict, the chondrules drifted sufficiently slowly that
the system only experienced condensation, i.e.~that at all time $P_K \ge P_{K,sat}$, where
$P_K$ is K's actual partial partial pressure.

We can follow the results of \cite{2004GeCoA..68.4971R}, invoking
a thermal time scale $t_T$ which measures the time for the chondrules to have cooled from being just cool enough
for meaningful recondensation to occur to temperatures sufficiently low that complete recondensation can be assumed.
 We assume that the gas and
chondrule fluid densities were high enough that Richter's $\tau_{cond} \ll t_\tau$;
and note that Richter's residence time $\tau_R$
is conceptually similar to our $\tdep$ (although we have the solid, not the gas phase, moving through space).
We will also consider the time $\tdif$ for K to diffusion through the chondrules near its condensation temperature.

To avoid generating an isotopic signature we need at least one of:
\EQ
\tdif \ll \tdep,
\EN
in which case the chondrules were in continuous equilibrium with the gas; or
\EQ
t_T \ll \tdep
\EN
in which case all the available potassium would have recondensed at once, avoiding any isotopic signatures. In this latter case
however, if
\EQ
t_T \ll \tdif,
\EN
we would expect the deposition of a potassium layer on the surface of the chondrules, and significant zoning within the chondrules.
That is significant because when chondrule precursors melted,
they lost potassium proportional to their mass, while when K recondensed it did so at a rate proportional
to the surface area: large grains would have become depleted compared to small grains.

\subsection{Alkali diffusion times}

In general, for element $i$,
\EQ
t_{\text{diff},i} = \frac{a^2}{\DDD_i},
\EN
where $\DDD_i$ is the appropriate diffusion coefficient for that element. Unfortunately, diffusion coefficient measurements for K and Na through
different minerals differ by orders of magnitude depending on the mineral.
In Table \ref{diff_coeffs} we list the diffusion coefficients for K and Na in rhyolite and albite \citep[][and references therein]{Brady2013},
as well as the associated diffusion timescales through $a=0.25$\,mm spheres.
We can see that, in general, Na diffuses rapidly and is expected to satisfy $\tdif \ll \tdep$, 
but, depending on the minerals involved, K may not.

\begin{table}
\caption{Diffusion parameters}
\centering
\begin{tabular}{l l l}
\hline \\  [-2ex]
Mineral & Na & K \\
\hline \\ [-2ex]
$\DDD_{\text{rhyolite}}$ & $\simeq 10^{-6} \cmcms$ &$\simeq 10^{-8} \cmcms$ \\
$t_{\text{diff, rhyolite}}$ & $\simeq 1.2 \times 10^3$\,s & $\simeq 6 \times 10^4$\,s \\
$t_{\text{diff, rhyolite}}$ & $\simeq 10^{-5} \tR^{-3/2}\,\Orb$ & $\simeq 5 \times 10^{-4} \tR^{-3/2}\,\Orb$ \\
\hline
$\DDD_{\text{albite}}$ & $\simeq 10^{-10} \cmcms$ & $\simeq 10^{-13} \cmcms$ \\
$t_{\text{diff, albite}}$ & $\simeq  7.6 \times 10^6$\,s & $ \simeq 8 \times 10^9$\,s \\
$t_{\text{diff, albite}}$ & $ \simeq 6.4 \times 10^{-2} \tR^{-3/2}\,\Orb$ & $ \simeq 66 \tR^{-3/2}\,\Orb$\\
\hline
\end{tabular}
\label{diff_coeffs}
\end{table}

This means that the lack of Na zoning seen by \cite{2008Sci...320.1617A}
is not unexpected for a locally generated depletion/enhancement of Na (which would be expected
to track that of K given their similar volatility): chondrules are expected to internally equilibrate Na on time scales
$t_\text{diff} \ll \tdep$.
The ``a few'' orders of magnitude difference between Na and K diffusivities also suggests that any observed difference in K and Na zoning
within a given chondrule would, if combined with a detailed internal diffusion model for that chondrule, provide a direct
measure for $\tdep$: the condensation temperatures for Na and K are sufficiently close that the two elements experienced similar values
of $\tdep$ and $t_T$.

\section{Conclusions}

The observed strongly varying degrees of potassium depletion between chondrules required the existence
of separate reservoirs with different alkali metal abundances.
We have explored how such different reservoirs could have been locally created in the solar nebula, concluding
that while they might have been straightforwardly produced by heating the gas enough to evaporate the alkalis and subsequently
transporting the solids through the gas, leaving the alkalis behind, the
time scales required were significant: sub-orbital to hundreds of orbits. Further, the lower end of the time scales are only available
if chondrule formation occurred at altitude, and chondrule precursors were very porous. 
This porosity requirement is problematic because high dust porosities are not expected to survive temperatures $T\sim 1000$\,K (far below
chondrule processing temperatures): silicate grains are expected to rapidly
coalesce at such temperatures, reducing their porosity \citep{2015Icar..254...56H}.
If chondrule formation occurred at the midplane,
the minimum time scales balloon to tens of orbits.
This is not entirely surprising: chondrules
are very small, and hence were very well tied to the gas \citep{2012Icar..220..162J,2015Icar..245...32H}. Moving chondrules through the gas was hence
a laborious process.

Such long time scales are in significant tension with
laboratory estimates of chondrule cooling rates, which were much faster. That tension would be resolved
if newly formed chondrules rapidly cooled enough to lock in their petrology \citep[$T \gtrsim 1400$\,K,][]{1990Metic..25..309H}, but subsequently
stayed warm ($T\gtrsim 1000$\,K) for long enough for them to have slipped through the gas. On the other hand,
long timescales could simplify the puzzling lack of Na zoning in chondrules seen by \cite{2008Sci...320.1617A}: if the chondrules stayed
warm long enough for Na (and possibly K) to have fully diffused through its host, then there would have been no need
for extreme concentrations of solids in chondrule forming regions. Staying at $T\gtrsim 1000$\,K for prolonged periods is significant
because that temperature is associated with thermal ionization of alkali metals which are significantly evaporated, allowing the MRI to be active \citep{1996ApJ...457..355G}.
This would support the case for MRI turbulent dissipation as a chondrule heating mechanism
\citep{2012ApJ...761...58H,2013ApJ...767L...2M,2014ApJ...791...62M}. 

If, however, such time scales spent warm were not available, then the origin of the different
reservoirs with differing elemental, but not isotopic, abundances becomes a major logistical problem. The lack of a potassium
isotopic signature rules out a pre-solar source for the different reservoirs, and any other model will have to wrestle with the
improbabilities associated with generating many reservoirs. That difficulty is compounded by complimentarity,
which argues that the reservoirs from which chondrules in a given chondrite formed were correlated.
Further, the presence of cold matrix strongly constrains the amount of thermal processing
that could have occurred in the solar nebula: most of the solids
were never thermally processed enough to evaporate potassium \citep{2015Icar..245...32H}, so if the potassium abundance variations
were thermal in origin, and were generated in the solar nebula, they must have been correlated with chondrule formation, lest
the thermally processed material show up in the matrix.

The long time scales required to have generated potassium abundance variations in the solar nebula can also
explain the lack of potassium isotope signatures throughout the Solar System, even between bodies with
wildly varying potassium abundances \citep{1995Metic..30Q.522H}. 
Isotopic equilibration is to be expected if, as in the case of chondrules, the time scales required to physically move the
solids through gas (which is required to deplete or enhance the potassium abundances) are long compared to K's evaporation
or recondensation time scales. Indeed, a potassium isotopic signature would
be of great use in determining the time scale of physical processes in the solar nebula.

\section*{Acknowledgements}
The research leading to these results received funding from NASA OSS grant NNX14AJ56G.

\bibliographystyle{yahapj}
\bibliography{paper.bib}

\end{document}